\documentclass[10pt,a4paper]{article}

\usepackage{epsfig}
\usepackage{amsmath}
\usepackage{amssymb}
\usepackage{verbatim}
\usepackage{bm}
\usepackage{dsfont}
\usepackage[footnotesize]{caption}
\usepackage{subfigure}
\usepackage{cancel}
\usepackage{cite}
\usepackage{textcomp}
\usepackage{calc}
\usepackage{geometry}
\usepackage{bbm}
\usepackage{xcolor}
\usepackage{fancyhdr}
\usepackage{slashed}
\usepackage{lipsum} 
\usepackage[nottoc,numbib]{tocbibind}
\usepackage{xfrac}

\usepackage{hyperref}
\hypersetup{colorlinks=true,urlcolor=blue,linkcolor=red,citecolor=teal}

\renewcommand{\baselinestretch}{1.5}

\begin{document}


\setlength{\headheight}{13.6pt}
\pagestyle{fancyplain}
\lhead{\fancyplain{}{Modern Cosmology, an Amuse-Gueule}}
\rhead{\fancyplain{}{Kai Schmitz}}


\title{\vspace{-1.5cm}Modern Cosmology, an Amuse-Gueule}

\author{Kai Schmitz%
\footnote{Theoretical Physics Department, CERN, 1211 Geneva 23, Switzerland. Email: \href{mailto:kai.schmitz@cern.ch}{kai.schmitz@cern.ch}}}

\date{}

\maketitle

\centerline{\textbf{Abstract}}

This essay is a nontechnical primer for a broader audience, in which I paint a broad-brush picture of modern cosmology.
I begin by reviewing the evidence for the big bang, including the expansion of our Universe, the cosmic microwave background, and the primordial abundances of the light elements.
Next, I discuss how these and other cosmological observations can be well explained by means of the concordance model of cosmology, putting a particular emphasis on the composition of the cosmic energy budget in terms of visible matter, dark matter, and dark energy.
This sets the stage for a short overview of the history of the Universe from the earliest moments of its existence all the way to the accelerated expansion at late times and beyond.
Finally, I summarize the current status of the field, including the challenges it is currently facing such as the Hubble tension, and conclude with an outlook onto the bright future that awaits us in the coming years and decades.
The text is complemented by an extensive bibliography serving as a guide for readers who wish to delve deeper.




{\hypersetup{linkcolor=black}\renewcommand{\baselinestretch}{1} \tableofcontents}


\newpage
\section{Evidence for the big bang}


\subsection{Hubble expansion}


Modern cosmology~\cite{Kolb:1990vq,Mukhanov:2005sc,Weinberg:2008zzc,Dodelson:2003ft,Peebles:2020aaa} witnessed its dawn about a hundred years ago, in the twenties of the last century, shortly after Albert Einstein had laid the necessary theoretical groundwork in the form of his general theory of relativity~\cite{Misner:1973prb,Carroll:2004st}.
New astronomical observations in 1920s revealed that the spiral nebulae scattered across the night sky are in fact distant galaxies, implying that our Universe is vastly larger than previously thought and our home galaxy, the Milky Way, just one among nearly countless many.
This realization lends support to the Copernican principle, the idea that we do not observe the Universe from a privileged vantage point.
Together with the observation that the Universe appears to look the same no matter where we point our telescopes, the Copernican principle provides the basis for the cosmological principle: 
On distance scales larger than galaxy superclusters, the Universe is homogeneous and isotropic; its properties are the same everywhere and in all directions.


General relativity in combination with the cosmological principle almost unavoidably predicts that the Universe must be expanding (or contracting, which, however, is not supported by observations).  
This theoretical prediction was first derived by Alexander Friedmann in 1922~\cite{Friedmann:1922aa}, who noticed that all physical distances in a spatially homogeneous and isotropic Universe must grow in proportion to a universal factor $R\left(t\right)$, the cosmic scale factor~\cite{Barnes:2022gjb}, which is independent of location and only depends on the cosmic time coordinate $t$.
The time evolution of this scale factor is described by two equations that Friedmann derived from Einstein's field equations and which are now known as the Friedmann equations.
Five years later, Georges Lema\^itre, not aware of Friedmann's work, independently argued that the Universe might be expanding as well as that an expanding Universe must have a finite age~\cite{Lemaitre:1927zz}:
If the physical distance between two points in space is constantly growing, the two points must have been closer together in the past, until, after rewinding the cosmic clock by a sufficient amount, their separation approaches zero. 
Lema\^itre referred to his theory as the hypothesis of the primeval atom, which would later develop into the idea of the big bang (a term coined by Fred Hoyle in 1949 in a BBC radio interview). 


According to the big-bang model, the Universe originated from an extremely hot and dense initial state\,---\,the further back in time, the hotter and denser\,---\,followed by a stage of decelerating expansion causing it to cool down and become more and more dilute.
The extrapolation back in time that underlies the big-bang model ultimately breaks down at $t=0$, where the scale factor goes to zero, $R\rightarrow0$, and the density and temperature of the Universe become infinitely large. 
The term \textit{big bang} is sometimes used to denote this singularity, even though it merely signals that we have pushed our theory beyond its range of validity.
After all, big-bang cosmology is based on general relativity, a classical theory of gravitation, and can therefore be at most trusted up to the Planck era at $t \sim 10^{-43}\,\textrm{s}$.
At earlier times, quantum-gravity effects are expected to become relevant and regulate the big-bang singularity, albeit in absence of a fully developed theory of quantum gravity, only highly speculative statements are possible at present.
In any case, it is clear that the scale factor never vanishes during the stages of the cosmic expansion history that can be described by the standard big-bang model.
An infinitely large Universe is therefore always infinitely large in big-bang cosmology.
The big bang had in particular no center or spatial origin but occurred everywhere in space simultaneously. 


In addition to his theoretical work on the earliest moments of the Universe, Lema\^itre also pointed out how the expansion of the Universe would manifest itself in astronomical observations.
He realized that comoving observers in an expanding Universe would mutually recede from each other, just as raisins in a rising yeast dough recede from each other, even though they do not move with respect to the dough.
In particular, he computed that the expected recessional velocities should be linearly proportional to the physical distances between the observers. 
In the following years, Edwin Hubble was able to observationally confirm this expectation~\cite{Hubble:1929ig}, in collaboration with Milton Humason~\cite{Hubble:1931zz} and building on earlier work by Vesto Slipher~\cite{Slipher:1913aa} and Knut Lundmark~\cite{Lundmark1924aa}. 
To this end, Hubble determined the distances $D$ to a handful of galaxies in our cosmic neighbourhood as well as their recessional velocities $v$ (by measuring the redshift of the light received from these galaxies), which allowed him to identify a universal linear relation of the form $v \approx H_0 D$.
This relation agrees with Lema\^itre's prediction, provides key evidence for the expansion of the Universe and the big bang, and is now known as the Hubble--Lema\^itre law.
The constant of proportionality in the Hubble--Lema\^itre law is the Hubble constant, $H_0 \approx 70\,\textrm{km}/\textrm{s}/\textrm{Mpc}$.
It corresponds to the present-day value of the Hubble rate $H = \dot{R}/R$, which is defined in terms of the scale factor $R$ and its time derivative $\dot{R}$ and which was much larger than $H_0$ in the early Universe.
The numerical value of $H_0$ tell us that for each megaparsec that a galaxy is further away from us, its recessional velocity increases by roughly $70\,\textrm{km}/\textrm{s}$.
Moreover, one can show that its inverse value is closely related to the age of the Universe of about 13.8 billion years.
$H_0$ has always been and still is one of the most important observables in modern cosmology; its precise value is the subject of ongoing research, as we will discuss at the end of this essay.


\subsection{Cosmic microwave background}
\label{subsec:cmb}


Olbers' paradox confronts us with the puzzle of why the night sky is dark\,---\,if the Universe were static as well as large and old enough, we would actually expect it to be as bright as the surface of the sun~\cite{Harrison:1987aaa}.
Big-bang cosmology resolves this paradox. 
To see this, we first note that the big-bang model does in fact imply a radiation background that is constantly reaching us from all directions in the sky, a theoretical prediction that was first made by Ralph Alpher, George Gamow, and Robert Herman in the late 1940s~\cite{Gamow:1948pob,Alpher:1948srz,Alpher:1948gsu}.
This relic radiation can be thought of as the afterglow of the big bang and is an immediate consequence of the dropping temperature in the early Universe.
Roughly 380,000 years into its existence, the Universe had sufficiently cooled down, such that electrons and protons could begin to form neutral hydrogen.
Following this process called recombination~\cite{Peebles:1968ja}, photons ceased to undergo Thomson scattering with free electrons, which allowed them to decouple from the hydrogen gas and henceforth propagate freely through the Universe.
Photon decoupling thus gave rise to what is known as the surface of last scattering, a two-dimensional sphere with the solar system at its center that encompasses all spatial locations from which relic photons had a chance to travel to us during the last 13.8 billion years after they had last scattered with an electron. 
Even though the Universe as a whole is probably infinitely large, this surface defines the size of the \textit{observable} Universe, a horizon beyond which we cannot see because of the finite age of the Universe and the finite speed of light. 
Of course, the Universe has kept expanding since photon decoupling, which is why this horizon is currently located at a distance of about 45 rather than 13.8 billion light years.
Besides, the cosmic expansion is also the key to resolving Olbers' paradox.
During their cosmic journey towards us, the relic photons are subject to cosmological redshift, \textit{i.e.}, their wavelength becomes stretched because of the cosmic expansion by around a factor 1100, such that they reach us today in the form of microwave radiation. 
This cosmic microwave background (CMB) radiation~\cite{Durrer:2020fza}, while a fundamental prediction of big-bang cosmology, is not visible to the naked eye, which explains why the night sky is dark.


In 1965, less than two decades after it had been theoretically predicted, Arno Penzias and Robert Wilson accidentally detected the CMB in observational data taken with the Holmdel Horn Antenna in New Jersey~\cite{Penzias:1965wn}.
This breakthrough marked another triumph of big-bang cosmology, which was crucial in establishing it as the basis of modern cosmology.
As expected for a photon gas of thermal origin, the specific intensity of the CMB agrees perfectly with the spectrum of a blackbody, a blackbody at a temperature of around three kelvin, which is why the CMB is also referred to as $3\,\textrm{K}$ radiation.
In fact, the CMB spectrum recorded by NASA's Cosmic Background Explorer (COBE) satellite mission in the early 1990s represents the best blackbody spectrum ever measured~\cite{Fixsen:1996nj}.
The CMB therefore provides striking evidence for one of the most important predictions of big-bang cosmology, namely, that the Universe must have been in a hot thermal state in its past.

\begin{figure}

\centering
\includegraphics[width=0.89\textwidth]{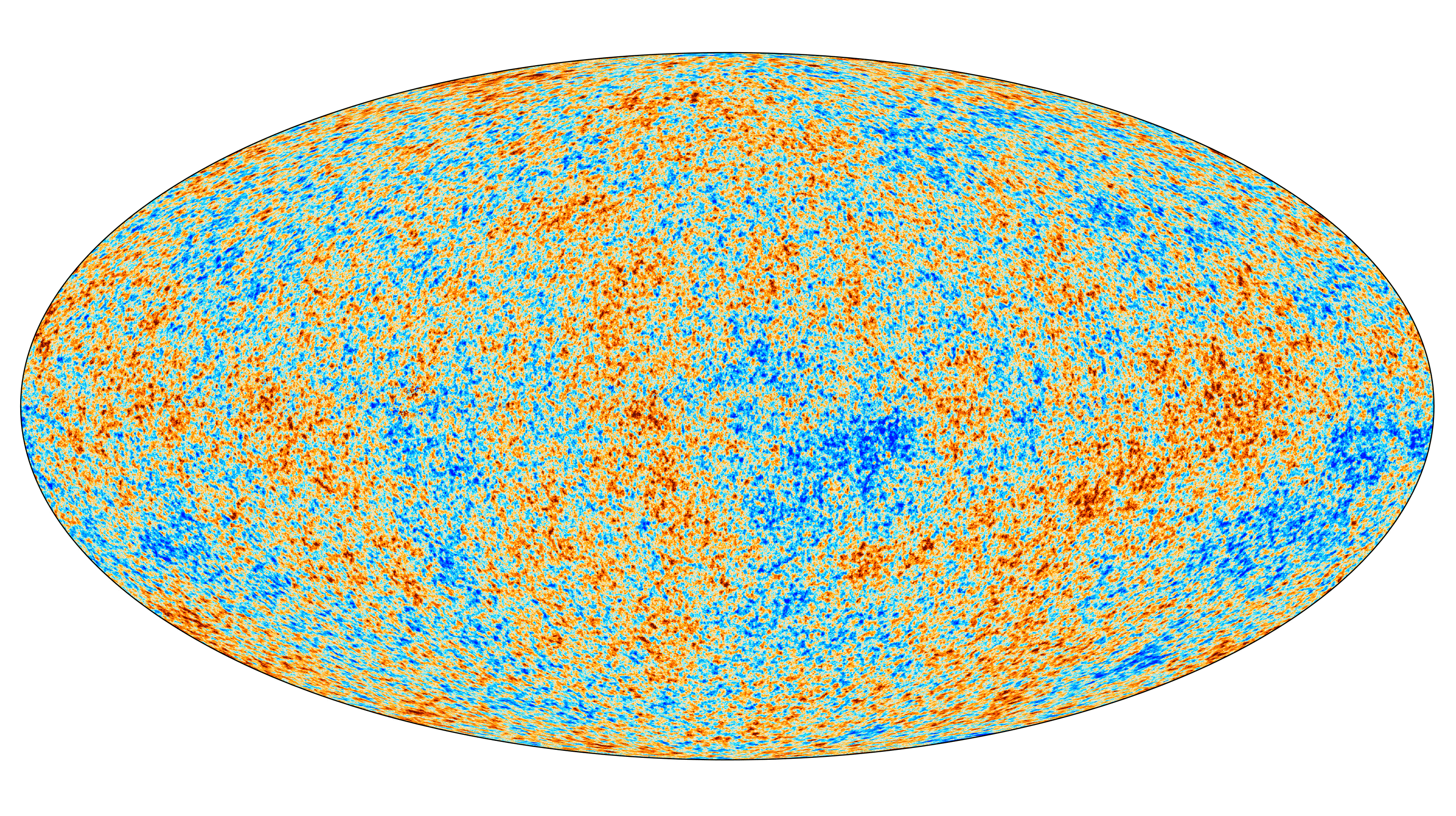} 

\caption{Baby picture of our Universe roughly 380,000 years after the big bang: temperature anisotropies in the CMB as seen by ESA's PLANCK satellite mission~\cite{Planck:2018nkj}. The mean CMB temperature across the entire sky is $T_{\rm CMB} = 2.72548 \pm 0.00057\,\textrm{K}$~\cite{Fixsen:2009ug}; the red spots are hotter than the blue spots by a few hundred microkelvin.
Image credit: ESA and the PLANCK Collaboration.}

\label{fig:cmb}

\end{figure}

At first sight, the temperature of the CMB blackbody spectrum appears to be completely isotropic across the sky, in accord with the cosmological principle.
Upon closer inspection, however, it turns out that it exhibits minuscule fluctuations around its mean value of around three kelvin.
These temperature anisotropies are the consequence of density perturbations in the baryon--photon fluid at the time of recombination, \textit{i.e.}, tiny over- and underdensities that are bound to develop into the large-scale structure of the Universe at later times.
Big-bang cosmology does not provide an explanation for the origin of these primordial seeds of structure formation.
They are instead attributed to the initial conditions of the hot big bang, which means that their generation must be addressed in theories like inflationary cosmology or one of its alternative (see Sec.~\ref{subsec:ics}). 
The most prominent contribution to the CMB temperature anisotropies has an amplitude of a few millikelvin and the shape of a dipole. 
The standard view is to ascribe this contribution to our relative motion with respect to the CMB rest frame, even though there is an on-going debate as to whether at least parts of it may not have a different explanation that is not of kinematic origin~\cite{Secrest:2020has,Perivolaropoulos:2021jda}.
If the dipole contribution is subtracted from the CMB temperature map, one is left with anisotropies of the order of a few hundred microkelvin; see Fig.~\ref{fig:cmb}. 


Another important feature of the CMB is that it is linearly polarized because of Thomson scattering around the time of photon decoupling. 
Similarly to its temperature $T$, the polarization of the CMB also exhibits anisotropies across the sky.
These polarization anisotropies are typically decomposed into two contributions: a curl-less $E$ mode and a divergence-free $B$ mode. 
Together, the auto and cross correlation power spectra of these different types of anisotropies ($TT$, $TE$, $EE$, $BB$) encode a wealth of information about the early Universe and the propagation of the CMB photons towards us.
Here, temperature and $E$-mode polarization anisotropies can be sourced by two different types of primordial perturbations: density fluctuations (\textit{i.e.}, scalar perturbations) and primordial gravitational waves (\textit{i.e.}, tensor perturbations). 
Large-scale $B$-mode polarization, however, can only be generated by tensor perturbations (on small scales, gravitational lensing converts $E$-mode into $B$-mode polarization). 
Past CMB experiments, including the COBE~\cite{Fixsen:1996nj}, WMAP~\cite{WMAP:2012fli}, and PLANCK~\cite{Planck:2018nkj} space missions, did unfortunately not succeed in finding evidence for such a $B$-mode signal.
Hope therefore rests on future CMB polarization experiments  such as BICEP Array~\cite{Hui:2018cvg}, CMB-S4~\cite{CMB-S4:2016ple}, LiteBIRD~\cite{LiteBIRD:2022cnt}, PICO~\cite{NASAPICO:2019thw}, and the Simons Observatory~\cite{SimonsObservatory:2018koc}.
The detection of a large-scale $B$-mode polarization signal by any of these experiments would amount to the discovery of primordial gravitational waves (GWs), a breakthrough in early-Universe cosmology.


In addition to primordial tensor perturbations, the CMB allows one to measure or constrain a large range of further observables.
Another important example is the spectral tilt of the primordial scalar power spectrum, $n_s$, which is a measure for the scale dependence of the primordial density fluctuations in the baryon--photon fluid.
Its value is now known to such a precision that a scale-invariant origin of these fluctuations appears extremely unlikely~\cite{Planck:2018vyg}.
In other words, whatever physics set the initial conditions of the big bang, it must have involved some scale-dependent dynamical process (such as, \textit{e.g.}, cosmic inflation in the so-called slow-roll approximation, which slightly breaks the scale invariance of perfect de Sitter space).
Furthermore, the CMB contains information on the Hubble constant, the spatial curvature of the Universe, its energy budget, and many other quantities such as primordial non-Gaussianities and isocurvature perturbations.
A more detailed discussion of these aspects of CMB physics is unfortunately beyond the scope of this essay.
Instead, we conclude by emphasizing the importance of the CMB as a key driver of modern cosmology. 
It is a treasure chest for cosmologists that will continue to play an essential role in the field for many decades to come.


\subsection{Primordial nucleosynthesis}
\label{subsec:bbn}


The CMB represents an opaque veil that we cannot see beyond.
It is, however, not our earliest probe of the early Universe, which brings us back to the work of George Gamow and his collaborators, whose prediction of the CMB resulted in fact from their interest in a different question, namely, the origin of the chemical elements.
Gamow et al.\ were particularly intrigued by the idea of the big bang because it provides the right conditions for the synthesis of light nuclei from a hot, but gradually cooling plasma of free protons and neutrons. 
Their seminal work~\cite{Alpher:1948ve} laid the foundation of what is now known as primordial or big-bang nucleosynthesis (BBN)~\cite{Cyburt:2015mya,Pitrou:2018cgg} and which provides us with another window onto the early Universe. 
Although BBN is a slightly more indirect probe of the big bang compared to the CMB, it allows us to rewind the cosmic clock all the way back to the first second on the cosmic time axis, \textit{i.e.}, a time when the temperature of the Universe was around one megaelectronvolt and hence a million times larger than during photon decoupling.


BBN theory predicts the primordial abundances of isotopes that were formed during the first 20 minutes or so, notably, deuterium, helium-3, helium-4, and lithium-7.
These predictions can be compared to astronomical measurements of the primordial element abundances.
Primordial deuterium leaves an imprint in the spectra of high-redshift quasar absorption systems; primordial helium-4 is found in low-metallicity regions of ionized hydrogen; and primordial lithium-7 shows up in the spectra of old metal-poor dwarf stars in the halo of our galaxy.
Apart from a slight discrepancy with respect to the lithium abundance~\cite{Fields:2011zzb}, which requires further study, one finds very good agreement between theory and observations.
This is a remarkable result, especially, in view of the fact that the measured abundances range over nine orders of magnitude. 
The success of BBN theory thus represents another triumph of big-bang cosmology. 
Moreover, it leads us to the conclusion that the laws of particle and nuclear physics, which we can test in terrestrial laboratories and which are a crucial input to BBN theory, must have also applied in the early Universe. 
This is a profound insight into the nature of the physical world.
Had we to give up on the universality of the laws of physics, it would become much harder, if not impossible, to make meaningful statements about the cosmology of the early Universe.


Among the most important input for BBN theory are the cross sections for nuclear interaction rates.
Progress in the field therefore crucially depends on improved measurements of these cross sections in laboratories, such as, \textit{e.g.}, the recent measurement of the rate of deuterium burning by the LUNA collaboration~\cite{Mossa:2020gjc}.
Once all nuclear cross sections are set to their measured values, BBN more or less turns into a one-parameter theory, at least as long as one does not introduce any new physics beyond the Standard Model (SM) of elementary particle physics. 
The only remaining free parameter is then the baryon number density $n_b$ or, equivalently, the baryon-to-photon ratio $\eta_b = n_b/n_\gamma$ at the time of nucleosynthesis.
$\eta_b$ is an important cosmological parameter. 
It also affects the properties of the baryon--photon fluid at the time of photon decoupling and can hence be deduced from CMB observations.
As we find ourselves living in a Universe that does not contain any appreciable amount of antimatter, $\eta_b$ can also be regarded as a measure of the matter--antimatter asymmetry, or baryon asymmetry, of the Universe, $\eta_b = \left(n_b - n_{\bar b}\right)/n_\gamma$, where $n_{\bar b} \approx 0$ denotes the present-day number density of antibaryons.
BBN theory is able to reproduce the measured abundances of the light elements for $\eta_b \approx 6 \times 10^{-10}$~\cite{ParticleDataGroup:2020ssz}.
Remarkably enough, this value is in excellent agreement with the value inferred from the CMB~\cite{Planck:2018vyg}.


Among all of its achievements, this concordance between BBN and the CMB is arguably the biggest success of the big-bang model.
It is furthermore an amazing confirmation that, based on the methods of modern cosmology, we are indeed able to reconstruct the history of our Universe all the way back to the first second of its existence. 
Despite the countless open questions that remain to be answered in cosmology, it is therefore fair to say that \textit{the big bang did indeed happen}\,---\,in the sense that, 13.8 billion years ago, the Universe was indeed filled by a rapidly expanding hot thermal plasma that gradually became cooler and more and more dilute while giving birth to all of the visible matter that permeates the cosmos today.
Certainly, the big-bang model cannot be the end of the story.
Ultimately, it will need to be embedded into a more fundamental theory that also explains the origin and initial conditions of the big bang, probably in the context of quantum gravity (see Sec.~\ref{subsec:ics}).
However, just like Newton's theory of gravitation will always remain embedded in Einstein's theory of general relativity, the big bang will always remain embedded in the more fundamental theory that is going to succeed it.  


The success of BBN theory, next to its important implications for our worldview, also represents a powerful tool for particle cosmology.
It enables us, \textit{e.g.}, to constrain the properties of new subatomic particles that may exist beyond the Standard Model (BSM). 
This includes BSM models that predict the existence of new massless or very light particles, so-called dark radiation~\cite{Archidiacono:2013fha}, which can affect the speed of the Hubble expansion during nucleosynthesis.
The amount of dark radiation is typically quantified in terms of a quantity called $N_{\rm eff}$, the effective number of relativistic neutrino species, which counts new radiation degrees of freedom as if they corresponded to new generations of neutrinos.
$N_{\rm eff}$ is another important cosmological observable that can be constrained by BBN and CMB observations.
With a bit of luck, its precise value may hold the key to the discovery of new particles.
Besides that, the outcome of BBN is also sensitive to the presence of heavier particles that decay around or after nucleosynthesis~\cite{Kawasaki:2004qu}.
In some scenarios, such effects may be beneficial, \textit{e.g.}, because they help to ameliorate the lithium problem.
In most cases, however, they spoil the success of BBN theory, which then allows one to put bounds on the allowed mass and lifetime of the decaying particle.
Regardless of these details, though, it is clear that BBN offers a powerful test that all models of new physics must pass.


\section{Concordance model of cosmology}


\subsection{Visible matter}
\label{subsec:visible}


The excellent agreement between BBN and CMB observations represents an important pillar for the current concordance model of cosmology: the Lambda-cold-dark-matter ($\Lambda$CDM) model, which succeeds in explaining the bulk of all cosmological data available today.
At its core, the $\Lambda$CDM model is a particularly simple model, which, in its minimal form, features only six independent parameters.
This simplicity, however, does not come for free; the price to pay is a blatant discrepancy with the Standard Model of particle physics.
The crucial point is that the $\Lambda$CDM model describes the composition of our Universe in terms of three main ingredients\,---\,visible matter, dark matter, and dark energy\,---\,all of which call for BSM physics in one way or another.
In addition, it relies on initial conditions that cannot be explained solely based on the Standard Model in combination with Einstein gravity.
Cosmology therefore tells us that our understanding of the subatomic world is incomplete.
This profound conclusion establishes a deep connection between microscopic and macroscopic physics and serves as a strong motivation for new-physics searches in terrestrial high-energy experiments.


The name of the model, $\Lambda$CDM, refers to the fact that it assumes dark matter to be cold and dark energy to be described by a mere cosmological constant $\Lambda$; we will discuss both aspects in more detail below.
The minimal version of the model moreover assumes a flat spatial geometry, which is only possible if the total energy density of the Universe corresponds to a specific critical value, $\rho_{\rm crit} = 3 H_0^2 M_{\rm Pl}^2 \approx 10^{-26}\,\textrm{kg}/\textrm{m}^3$ (where $M_{\rm Pl}$ denotes the reduced Planck mass).
This critical density allows one to introduce the density parameters  $\Omega_b = \rho_b/\rho_{\rm crit}$, $\Omega_{\rm cdm} = \rho_{\rm cdm}/\rho_{\rm crit}$, and $\Omega_\Lambda = \rho_\Lambda/\rho_{\rm crit}$ for baryons, cold dark matter (CDM), and dark energy, respectively.
Omitting the negligibly small present-day energy densities of photons and neutrinos, these parameters satisfy by construction $\Omega_b + \Omega_{\rm cdm}  + \Omega_\Lambda = 1$ in a flat Universe.
Here, the energy densities of baryons and dark matter, $\rho_b \propto \Omega_b H_0^2$ and $\rho_{\rm cdm} \propto \Omega_{\rm cdm} H_0^2$, constitute two of the six free input parameters of the minimal $\Lambda$CDM model; the remaining four parameters being: (i) the age of the Universe, $t_0$; (ii) the spectral index of the primordial scalar power spectrum, $n_s$; (iii) the amplitude of the primordial scalar power spectrum, $A_s$; and (iv) the optical depth due to CMB photons undergoing Compton scattering during their cosmic journey towards us, $\tau$.


There is a wealth of observational techniques that allow one to measure or constrain the parameters of the $\Lambda$CDM model.
In addition to BBN and the CMB, we wish to highlight two of them: the galaxy power spectrum and the supernova Hubble diagram.
The former corresponds to the Fourier transform of the galaxy two-point correlation function and can be measured in galaxy surveys. 
It is an especially  useful cosmological probe because it exhibits a characteristic oscillatory feature, so-called baryonic acoustic oscillations (BAOs)~\cite{Percival:2007yw,SDSS:2009ocz}, which are related to the oscillations in the primordial baryon--photon fluid.
The distance scale on which these oscillations occur can be used as a \textit{standard ruler} that determines the size of the sound horizon at the time of photon decoupling. 
Meanwhile, the supernova Hubble diagram is constructed from measuring the redshift $z$ and distance $D$ of type-Ia supernovae (SNe)~\cite{SupernovaSearchTeam:1998fmf,SupernovaCosmologyProject:1998vns}.
At low redshift, these two quantities are linearly related to each other according to the Hubble--Lema\^itre law, $z\approx v = H_0 D$.
At larger redshift, however, the distance--redshift relation becomes nonlinear, with the exact relation being sensitive to the cosmological parameters.


\begin{figure}

\centering
\includegraphics[width=0.67\textwidth]{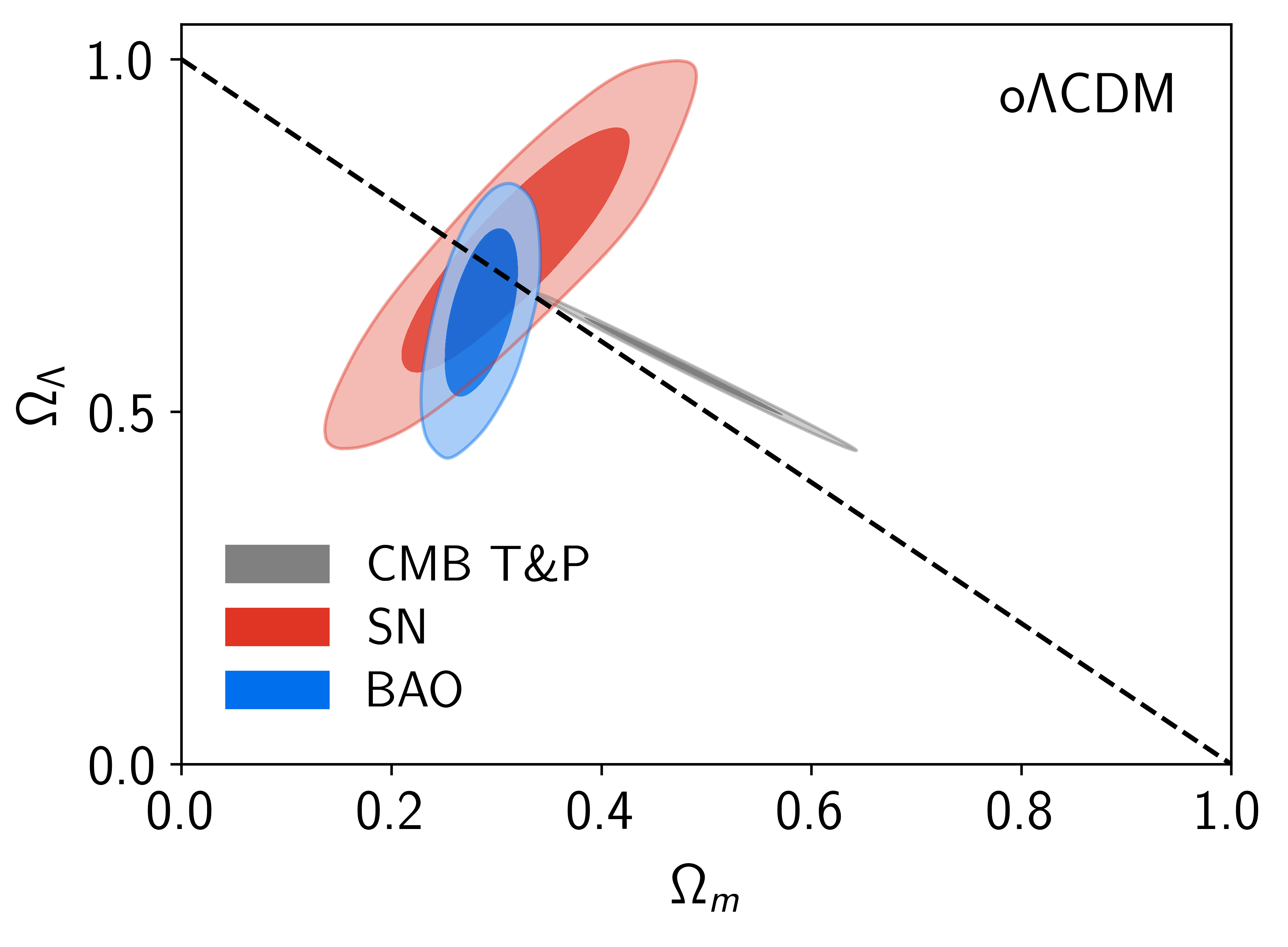} 

\caption{Constraints on the matter and dark-energy density parameters $\Omega_m$ and $\Omega_\Lambda$ based on CMB temperature (T) and polarization (P) data, type-Ia supernovae, and baryonic acoustic oscillations in the galaxy power spectrum~\cite{eBOSS:2020yzd}.
The dashed diagonal line corresponds to a flat Universe, $\Omega_m + \Omega_\Lambda = 1$.
The label ``o$\Lambda$CDM'' refers to an extension of the minimal six-parameter $\Lambda$CDM model that also allows for nonzero curvature (\textit{i.e.}, values of $\Omega_m + \Omega_\Lambda$ larger and smaller than 1).
Image credit: The eBOSS Collaboration.}

\label{fig:parameters}

\end{figure}


Together with the CMB temperature and polarization anisotropies, BAO and SNe observations allow for a precise determination of the matter and dark-energy density parameters, $\Omega_m = \Omega_b + \Omega_{\rm cdm}$ and $\Omega_\Lambda$; see Fig.~\ref{fig:parameters}.
The result of this measurement is that we live indeed in a flat Universe, $\Omega_m + \Omega_\Lambda \approx 1$, that is dominated by dark energy, $\Omega_\Lambda \approx 0.7$.
In the next step, we can combine this result with the measurement of the baryon density parameter from BBN and CMB observations, $\Omega_b \approx 0.05$, which leads us to the fascinating conclusion that the visible matter content of the Universe only accounts for around $5\,\%$ of its energy budget.
$95\,\%$ of the Universe is dark, with around $25\,\%$ of its energy contained in dark matter and $70\,\%$ of its energy made up of dark energy.


On the one hand, an energy fraction of $5\,\%$ in visible baryonic matter may not seem like much, especially when compared to the much larger energy densities of dark matter and dark energy.
On the other hand, it turns out to be a remarkably large value when compared to our naive theoretical expectation based on our understanding of the hot big bang.
Assuming matter--antimatter-symmetric initial conditions at very early times, we would expect that matter and antimatter almost completely annihilated each other in the early Universe, leaving behind nothing but radiation.
In a Universe where such an annihilation catastrophe takes place, the baryon-to-photon ratio would freeze out at a tiny value, suppressed by many orders of magnitude compared to the value that we actually measure in BBN and CMB observations of around $\eta_b \approx 6 \times 10^{-10}$. 
The fact that we do find a large baryon density compared to this theoretical expectation therefore indicates that the primordial plasma must have exhibited a baryon asymmetry at the time of baryon freeze-out.
Roughly speaking, for each billion antiparticles, there must have been a billion and one particle, such that after baryon--antibaryon annihilation the one excess baryon would survive in the plasma.
These surviving baryons now constitute the entire visible matter content of the present Universe, while there are no traces of primordial antimatter left.


The baryon asymmetry of the Universe (BAU) is one of the biggest unresolved problems in fundamental physics.
Naively, one may be tempted to attribute its origin simply to the initial conditions of the Universe.
This supposition, however, would not only be tantamount to a surrender, it would also be in conflict with the paradigm of cosmic inflation (see Sec.~\ref{subsec:ics}), according to which the hot big bang must have originated from a matter--antimatter-symmetric initial state.
It is therefore expected that the baryon asymmetry is the outcome of a dynamical process in the hot thermal phase of the early Universe.
This process, commonly referred to as baryogenesis, needs to satisfy three necessary conditions.
As noted by Andrei Sakharov in 1967~\cite{Sakharov:1967dj}, any successful baryogenesis scenario must necessarily (i) violate baryon number conservation, (ii) violate charge ($C$) and charge--parity ($CP$) invariance, and (iii) involve a departure from thermal equilibrium.
These conditions cannot be fulfilled in the Standard Model, which does not feature a sufficient amount of $CP$ violation and does not provide the basis for out-of-equilibrium interactions in the early Universe.
The BAU is thus not only a cosmological mystery, it also direct evidence for the existence of new BSM physics.
As such, it is the subject of intense on-going research efforts.
There is a vast number of baryogenesis models in the literature~\cite{Bodeker:2020ghk,DiBari:2021fhs}, among which the scenario of leptogenesis~\cite{Fukugita:1986hr,Buchmuller:2005eh} is arguably particularly promising.
Leptogenesis links the BAU to BSM physics in the neutrino sector and therefore leads to testable predictions for neutrino experiments.


\subsection{Dark matter}
\label{subsec:dm}


Dark matter~\cite{Bertone:2004pz,Bertone:2010zza} is roughly five times more abundant than ordinary matter, albeit much less understood. 
In fact, its very name is not only supposed to indicate that it does not interact with electromagnetic radiation, which is why we are not able to see it with our telescopes.
From this perspective, a better name would be \textit{transparent matter} anyway.
Instead, the name \textit{dark matter} also refers to our ignorance of its nature.
Just like the BAU, dark matter cannot be explained by the Standard Model and therefore provides evidence for new physics.

\begin{figure}[t]

\centering
\includegraphics[width=0.67\textwidth]{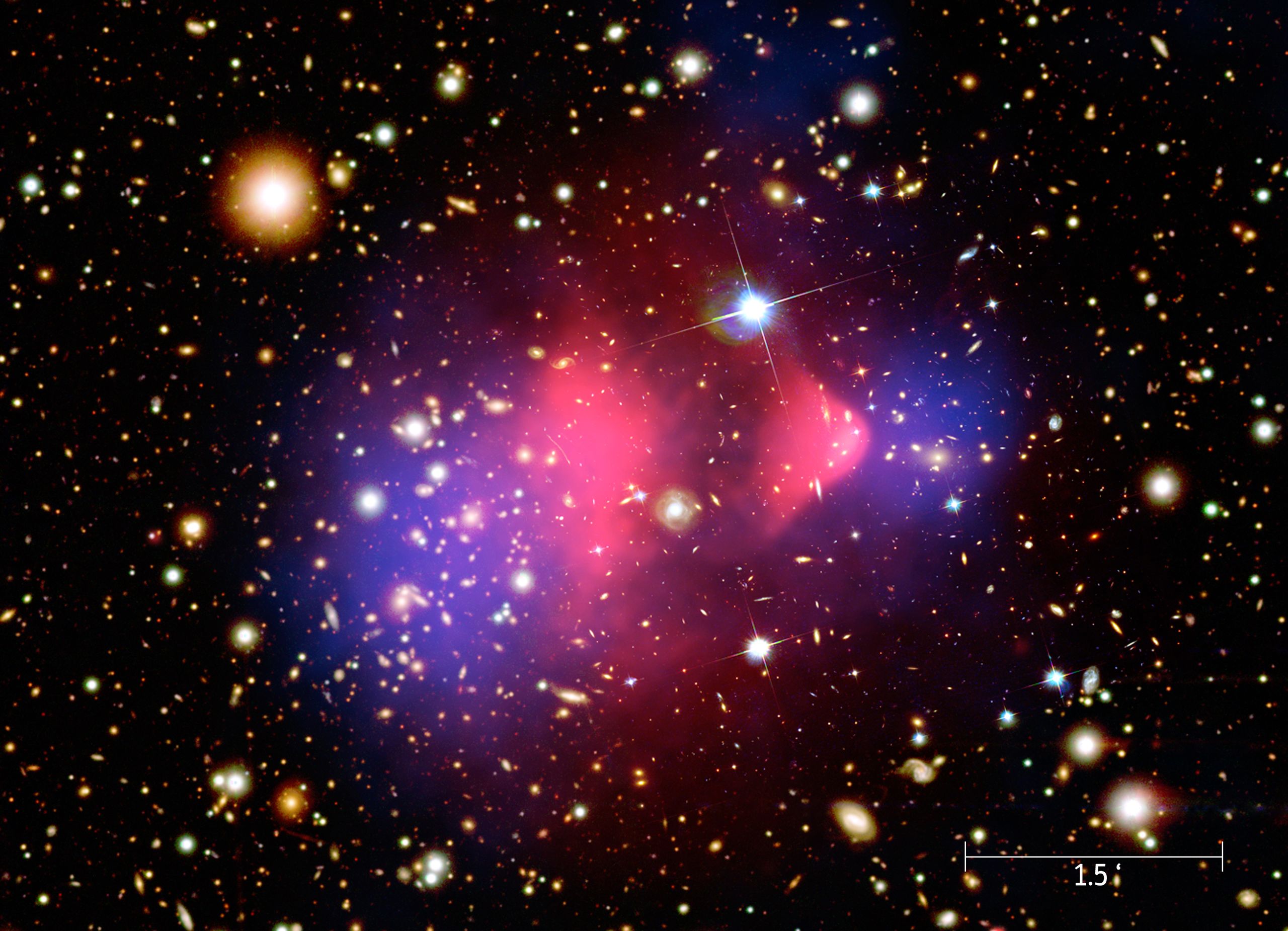} 

\caption{The Bullet Cluster, a collision of two galaxy clusters that provides compelling evidence for dark matter~\cite{Clowe:2006eq}. Most of the baryonic matter is contained in hot gas visible in X-rays (pink). Gravitational-lensing measurements (blue), however, indicate that most of the system's mass has already passed through the center, offsetting it from the hot gas. This can be explained by assuming that most of the system's mass belongs in fact to (nearly) collisionless nonbaryonic dark matter that does not slow down in consequence of the collision to the same extent as the dissipative gas component. At the same time, the luminous matter (galaxies) traces the gravitational potential induced by dark matter. In theories of modified gravity without dark matter, one would typically expect no offset between the baryonic matter and the mass distribution mapped by gravitational lensing. Image credit: NASA.}

\label{fig:bullet}

\end{figure}

As a gravitational phenomenon, the existence of dark matter is firmly established\,---\,in the sense that, across a large range of distance scales, the visible matter content of the Universe in combination with Einstein gravity does not suffice to explain our observations. 
We are therefore led to either assume (i) the presence of an invisible form of matter that only makes itself noticeable via its gravitational interaction with ordinary matter or (ii) a modification of general relativity~\cite{Milgrom:1983ca}.
The second scenario, however, while being successful in many cases, struggles to account for the entirety of existing observations (see Fig.~\ref{fig:bullet} for a famous example), which is why most cosmologists disfavor it.
Leaving aside the possibility of modified gravitational dynamics, the data then tells us that dark matter has been a key ingredient of the cosmic energy budget throughout cosmic history.


In the early Universe, dark matter is essential in setting the stage for structure formation.
Electrically neutral and nonbaryonic dark matter does not interact with the primordial baryon--photon fluid, which is why it is able to clump together and develop gravitational potential wells already much earlier than ordinary matter.
After photon decoupling, the baryonic matter then falls into these pre-existing potential walls, which triggers the formation of the large-scale structure that we observe in the Universe today.
This includes highly nonlinear structures such as galaxy halos and clusters, which could not exist if the early collapse of dark-matter overdensities had not provided the appropriate initial conditions.
Successful structure formation requires dark matter in particular to be cold, \textit{i.e.}, to have a nonrelativistic velocity at the onset of the matter-dominated era (see Secs.~\ref{subsec:rd} and \ref{subsec:structure}).
Cold dark matter results in a bottom-up scenario of structure formation, in which matter first collapses into smaller objects, which then gradually aggregate into larger objects. 
Alternatively, if dark matter were hot, \textit{i.e.}, relativistic at the onset of matter domination, structure would form in a top-down fashion, starting with the formation of large pancake-shaped objects, which then gradually fragment into smaller objects.
The three known neutrinos are an example of hot dark matter (as are light BSM neutrinos).
However, as cosmological observations and numerical simulations of structure formation strongly support the CDM paradigm, it is clear that hot dark matter can only contribute an insignificant fraction to the total dark-matter density.


In the late Universe, dark matter reveals its existence moreover in a wealth of observations that trace the total amount of mass in gravitationally collapsed structures.
This includes in particular the classic observations that historically led to the paradigm shift in favor of dark matter, such as the motion of galaxies in virialized galaxy clusters, which was pioneered by Fritz Zwicky in the 1930s~\cite{Zwicky:1933gu,Zwicky:1937zza}, and the rotation curves of spiral galaxies, which were first investigated by Vera Rubin and others in the 1960s and 1970s~\cite{Rubin:1970zza,Rubin:1980zd}. 
Today, these observations are complemented by analyses of gravitational lensing, redshift-space distortions, and many other probes.


It is therefore remarkable that, despite the overwhelming evidence for the existence of dark matter, we still do not know much about its nature at a fundamental level.
While we can list some of its key properties (nonbaryonic, cold, dark, etc.) with different levels of certainty, the mass of its fundamental building blocks, \textit{e.g.}, is still unknown. 
Dark matter may be composed of elementary scalar particles with a mass of the order of $10^{-22}\,\textrm{eV}$, a scenario that goes by the name of ultra-light or fuzzy dark matter~\cite{Hu:2000ke} and which is motivated by the so-called cusp--core problem of the $\Lambda$CDM model.
Meanwhile, it is also conceivable that a significant fraction of dark matter is made up of stellar-mass primordial black holes~\cite{Carr:2016drx}.
If some of these primordial black holes have masses a few times the mass of the sun, they might even be partially responsible for the binary black-hole merger events that have been detected by the LIGO and Virgo GW interferometers in recent years~\cite{Bird:2016dcv}.
The space of possibilities between these two extremes, fuzzy dark matter and primordial black holes, is filled by a sheer countless number of hypothetical scenarios for the nature and origin of dark matter.
In particular, there is no reason why dark matter should consist of only one component. 
Just like the SM sector, the dark-matter sector may well exhibit a nontrivial structure and encompass several ingredients.


Solving the enigma of dark matter represents a central challenge of 21$^{\rm st}$-century physics.
To achieve this goal, a vast array of experiments and observations is currently in operation or preparation:
Numerous laboratory experiments around the world strive to directly detect the exotic subatomic particles that may constitute our galaxy's dark-matter halo; telescopes and satellites look for indirect signatures of dark-matter annihilations or decays in outer space; and particle colliders attempt to directly produce dark-matter particles in high-energy collisions.
Other contributions to this book provide more details on the hunt for dark matter.


\subsection{Dark energy}
\label{subsec:de}


In a Universe filled only with visible and dark matter, one would expect the cosmic expansion to gradually slow down over time (\textit{i.e.}, a negative second time derivative of the scale factor, $\ddot{R}<0$).
Observations of type-Ia supernovae coordinated by Adam Riess, Brian Schmidt, and Saul Perlmutter in the 1990s, however, led to the astonishing conclusion that this type of behavior is not realized in our Universe. 
By extending the supernova Hubble diagram to large redshifts, the High-Z Supernova Search Team~\cite{SupernovaSearchTeam:1998fmf} and the Supernova Cosmology Project~\cite{SupernovaCosmologyProject:1998vns} were instead able to show that the cosmic expansion is currently accelerating (\textit{i.e.}, $\ddot{R}>0$).


In the framework of general relativity, this observation can only be explained if the stress--energy tensor on the right-hand side of the field equations receives a contribution that acts as a repulsive force.
For a perfect fluid with energy density $\rho$ and pressure $p$, this necessitates an equation of state $p = w\,\rho$ with $w < -1/3$, \textit{i.e.}, a negative pressure with large enough absolute value, $\left|p\right| > 1/3\,\rho$.
Dark energy is believed to be such an exotic form of energy with negative pressure, which permeates the cosmos and hence causes its accelerated expansion~\cite{Bamba:2012cp,Huterer:2017buf}.
In this sense, it represents another gravitational phenomenon that, next to dark matter, cannot be explained in terms of established physics. 
However, unlike dark matter, which is able to form gravitationally collapsed structures, dark energy is assumed to be largely spatially homogeneous across the Universe.


The concordance model of cosmology attributes the origin of dark energy to the energy density of the vacuum, \textit{i.e.}, empty space.
In $\Lambda$CDM cosmology, dark energy thus corresponds to an energy density constant in space and time (around 70\,\% of $\rho_{\rm crit} \approx 10^{-26}\,\textrm{kg}/\textrm{m}^3$) that serves as a cosmological constant $\Lambda$ in Einstein's field equations. 
This interpretation is motivated by the fact that vacuum energy possesses an equation of state given by $p=-\rho$, \textit{i.e.}, $w=-1$, which leads to accelerated expansion and agrees well with the cosmological data.
Indeed, a $\Lambda$CDM fit of the combined CMB, BAO, and SNe data fixes $w \approx -1$ within percent-level uncertainties~\cite{Planck:2018vyg}.


This phenomenological success is in stark contrast to our microscopic understanding of the cosmological constant $\Lambda$.
Historically, Einstein had first introduced a $\Lambda$ term in his field equations in an attempt to construct a cosmological solution that would describe a static Universe. 
Following the discovery of the expansion of the Universe by Hubble in the 1920s, he discarded this idea, referring to it as the greatest blunder of his life.
Thereafter, $\Lambda$ no longer attracted much interest for many decades.
Today, almost a quarter century after the discovery of the accelerated expansion, the situation is again very different and dark energy an active field of research.
Nonetheless, no satisfying theoretical explanation is currently in sight.
In principle, one may hope that quantum field theory (QFT) would allow one to estimate the size of the vacuum energy density based on the properties of the quantum vacuum. 
Estimates along these lines, however, turn out to be off by dozens of orders of magnitude, which represents a spectacular and maybe even embarrassing failure of our naive expectations. 


The cosmological constant appears in particular to be much \textit{smaller} than what one expects based on simple QFT arguments.
Part of the reason for this observation might be of anthropic nature. 
As famously pointed out by Steven Weinberg~\cite{Weinberg:1987dv,Weinberg:1988cp}, $\Lambda$ values not much larger than the one we observe would be inconsistent with the formation of complex structures in the Universe.
The accelerated expansion would simply set in too early during the expansion history of the Universe, leaving no time for the emergence of observers living in complex bound structures.
Similarly, the  small value of the cosmological constant results in what is known as the cosmological coincidence problem~\cite{Velten:2014nra}. 
Apparently, we currently live in a cosmological epoch where $\Omega_\Lambda$ and $\Omega_m$ are of the same order of magnitude and the vacuum-dominated stage of expansion has just begun.
Is this really just a coincidence, or is there a deeper reason why $\Omega_\Lambda$ must take a value that is neither much smaller nor much larger than $\Omega_m$?
In absence of compelling answers to these questions, the origin of the cosmological constant, or dark energy in general, remains one of the biggest puzzles in contemporary fundamental physics.


Alternative explanations of dark energy include, \textit{e.g.}, modified theories of gravity as well as extremely light scalar fields with cosmologically large de Broglie wavelength.
An interesting example of the latter scenario, commonly referred to as quintessence, are models that feature a parity-odd ``axion'' field coupling to photons.
In such axion quintessence models, dark energy can cause a rotation of the CMB polarization angle (\textit{i.e.}, cosmic birefringence).
Interestingly enough, first hints for such a signal have recently been reported, although more work is needed in order to better understand the contamination by astrophysical foregrounds~\cite{Minami:2020odp,Diego-Palazuelos:2022dsq}. 


A second property of quintessence is that its equation-of-state parameter can deviate from $w = -1$ and even vary as a function of time.
Thus, a clear measurement of $w\neq -1$ would indicate that dark energy is dynamical and rule out the standard scenario of a bare cosmological constant. 
This could have potentially severe implications for the ultimate fate of the Universe.
A cosmological constant characterized by $w=-1$ is expected to lead to an eternally lasting stage of accelerated expansion ($R\rightarrow\infty$, $H\rightarrow\textrm{const}$) that eventually results in a big freeze, \textit{i.e.}, a heat death slowly asymptoting towards zero temperature.
However, if dark energy should be accounted for by quintessence with a time-dependent equation of state, the future of the expanding Universe would be much more uncertain. 
In this case, dark energy may at some point even cease to dominate the expansion, which under certain conditions may lead to a big crunch, \textit{i.e.}, a recollapse into a singularity similar to the initial big-bang singularity ($R\rightarrow0$, $H\rightarrow\infty$).
Finally, one may consider the highly speculative idea of phantom dark energy, \textit{i.e.}, quintessence with negative kinetic energy and $w < -1$.
This scenario would result in a steadily increasing energy density of dark energy, which would cause a big rip, \textit{i.e.}, a singularity in the expansion rate ($R\rightarrow\infty$, $H\rightarrow\infty$) leading to the destruction of all bound objects, even at the subatomic level.
All of these ideas are of course very hard, if not impossible to test.
Still, future cosmological observations promise to shine more light on the equation of state of dark energy and hence its dynamical origin.


\section{History of the Universe}


\subsection{Initial conditions}
\label{subsec:ics}


In the previous section, we outlined the main building blocks of the $\Lambda$CDM concordance model of cosmology, which now sets the stage for an overview of the expansion history of the Universe during the last 13.8 billion years.
To this end, we first make a huge temporal leap from the ultimate fate of the Universe in the distant future back to the first moments of its existence in the distant past and then let cosmic time run forward again.


We begin our cosmic chronology by taking a closer look at the initial conditions of big-bang cosmology, which we already referred to several times to in our discussion but otherwise neglected thus far.
The crucial point is that these initial conditions, as became increasingly clear during the 1970s, are in fact far from generic.
Notably, the cosmological principle, one of the theoretical pillars of modern cosmology, only applies in big-bang cosmologies with highly fine-tuned initial conditions.
Picture, \textit{e.g.}, two CMB photons reaching us from diametrically opposite directions in the sky.
Both photons required 13.8 billion years for their cosmic journey towards us and thus stem from points on the surface of last scattering that, because nothing travels faster than a photon, never had a chance to exchange information with each other. 
Why does the CMB then nearly look the same in all directions? 
This question is at the core of the so-called \textit{horizon problem}, \textit{i.e.}, the puzzling realization that the early Universe must have been extremely homogeneous across a vast number of causally disconnected patches.
Similarly, the CMB and other cosmological probes indicate that our Universe is spatially flat to very good precision.
Based on the behavior of spatial curvature dictated by the Friedmann equations, one then concludes that our Universe must have been even flatter, in fact, exponentially close to perfect flatness in the past.
This is only possible if the energy density of the Universe is initially chosen to be extremely close to the critical energy density for spatial flatness, $\left|\rho - 3H^2M_{\rm Pl}^2\right| \lll 1$, which is by no means guaranteed and hence constitutes the \textit{flatness problem}.


A possible solution to these problems emerged around the early 1980s.
Alexei Starobinsky~\cite{Starobinsky:1979ty}, Alan Guth~\cite{Guth:1980zm}, Andrei Linde~\cite{Linde:1981mu}, and several other authors noted that a brief period of exponentially fast cosmic expansion could readily yield the required initial conditions for the hot big bang. 
According to this mechanism, dubbed \textit{cosmic inflation}~\cite{Baumann:2009ds} by Guth, the entire observable Universe was born from a single homogeneous and isotropic domain at early times, a causal patch of microscopic size whose volume was then blown up by a factor of around $10^{90}$ or even more within a timespan possibly as short as $10^{-36}$ seconds.
All parts of the observable Universe were therefore once in causal contact, which solves the horizon problem.
At the same time, any initial inhomogeneities and anisotropies were stretched out enormously, so that they lie beyond our cosmological horizon today.
The same also applies to hypothetical relics from even earlier times, including magnetic monopoles, \textit{i.e.}, one-dimensional topological defects in grand unified theories.
If produced during the hot big bang, monopoles would represent the most abundant form of matter in the Universe today, which clearly is not the case. 
Inflation solves this \textit{monopole problem}, which was actually one of the original motivations for the work of Guth et al., by pushing most, if not all relic monopoles beyond our horizon.
Similarly, spatial curvature is diluted during inflation, comparable to the surface curvature of a balloon, which decreases when the balloon is inflated.
This solves the flatness problem.


In view of these achievements, we conclude that the evolution of the classical spacetime background during inflation offers an attractive solution to the initial-conditions problems of big-bang cosmology.
But that is not all.
Remarkably enough, inflation also comes with a built-in mechanism for generating primordial density perturbations at the quantum level.
This is because the QFT description of inflation is based on the dynamics of a scalar \textit{inflaton} field (or several such fields). 
The evolution of the inflaton field background in a shallow scalar potential acts like a time-dependent cosmological constant and hence causes the exponential expansion, while its quantum fluctuations in combination with quantum fluctuations of the spacetime metric give rise to a spectrum of primordial scalar perturbations.
These perturbations are continuously produced during inflation from the quantum vacuum, whereupon they are stretched in length by the expansion, but not damped in amplitude.
Inflation thus explains the origin of the temperature fluctuations in the CMB and hence ultimately the origin of the large-scale structure of the Universe.
For each given QFT model, one is in particular able to predict two out of the six $\Lambda$CDM input parameters based on the microphysics of inflation: the amplitude and the spectral index of the primordial scalar power spectrum, $A_s$ and $n_s$.
The quantum fluctuations of the metric moreover give rise to primordial GWs across a large range of frequencies.
This primordial stochastic GW background can leave an imprint in the large-scale $B$-mode polarization of the CMB and be searched for directly in GW experiments.


We thus arrive at two remarkable conclusions:
First, inflation defines a new paradigm, \textit{inflationary cosmology}, which extends the old paradigm of big-bang cosmology by providing a compelling explanation of its initial conditions.
A central tenet of this paradigm is that all structure in our Universe is of quantum origin\,---\,generated during the first fractions of a second at extremely high energies, when the domain making up the observable Universe today was stretched from a microscopic size, maybe not much larger in diameter than a Planck length, $\ell_{\rm Pl} \approx 10^{-35}\,\textrm{m}$, to the size of a macroscopic object, say, an apple.
More details on the role of quantum mechanics in inflationary cosmology can be found in the chapter by Gabriele Veneziano. 
Second, we see that the theory of inflation is capable of making predictions that can be tested in CMB, GW, and other observations.


Next, let us turn to the theoretical description of inflation, which is currently to be considered as a general framework rather than a conclusive theory~\cite{Lyth:1998xn}.
The underlying idea of inflation is to use the potential energy density of the inflaton field as a form of vacuum energy that can drive a stage of accelerated expansion; however, the microscopic origin of this field remains unclear and leaves room for many speculations.
Indeed, many models of inflation, including the early models of old inflation~\cite{Guth:1980zm}, new inflation~\cite{Albrecht:1982wi}, chaotic inflation~\cite{Linde:1983gd}, etc., have the status of simple QFT toy models.
A lot of research during the last 40 years has therefore focused on the possible embedding of inflation into particle physics and string theory, with varying degrees of success.


On the particle physics side, it is, \textit{e.g.}, remarkable that even the SM Higgs field may play the role of the inflation, if it couples in a nonminimal way to the Ricci curvature scalar $\mathcal{R}$ in the gravitational action.
This model, known as Higgs inflation~\cite{Bezrukov:2007ep}, shares a lot of similarities with Starobinsky's original proposal, which extends the gravitational action by an $\mathcal{R}^2$ term~\cite{Starobinsky:1979ty}.
Currently, the predictions of Higgs and $\mathcal{R}^2$ inflation are in excellent agreement with the most recent CMB data.
Other field-theoretical models attempt to embed inflation into grand unification or establish a connection with the accelerated expansion of the current Universe.
The latter scenario is also known as quintessential inflation because it identifies the inflaton and quintessence as the same scalar field~\cite{deHaro:2021swo}.
This represents an economical scenario, even though there is no empirical reason to believe that the accelerated cosmic expansion at early and late times must be caused by the same dynamics.


Ultimately, inflation requires a quantum description of gravity, in order maintain control over gravitational corrections to the potential energy density of the inflaton field.
This motivates the construction of inflation models in the context of string theory~\cite{Baumann:2014nda}, which is one the main ambitions of the field of string cosmology~\cite{McAllister:2007bg}.
Inflation may even offer the possibility to test predictions of string theory by means of cosmological observations, which is otherwise a notoriously difficult task.
The vast landscape of possible vacuum states in string theory moreover lends support to the notion of the multiverse, \textit{i.e.}, the speculative idea that our Universe is just one among countless other universes that are continuously born during inflation, each possibly in a different vacuum state.
Anthropic arguments applied to inflation in the string landscape may notably explain the tiny value of the cosmological constant~\cite{Susskind:2003kw}, albeit such inferences, which most likely elude any possible means of experimental confirmation, are a source of controversy.
On the other hand, one may regard the multiverse as an unavoidable consequence of any stage of inflation that globally lasts forever but locally decays into causally disconnected bubbles. 
Inflation of this type is known as \textit{eternal inflation} and a generic prediction of many models~\cite{Vilenkin:1983xq}.


Much of the current quantum-gravity research on inflation focuses on its feasibility in view of the so-called \textit{swampland conjectures}, which attempt to delineate the boundary between effective field theories that do and that do not possess a consistent ultraviolet completion in quantum gravity, \textit{i.e.}, theories in the ``landscape'' and the ``swampland''~\cite{Palti:2019pca}.
Several of these conjectures pose a severe challenge to simple models of inflation~\cite{Agrawal:2018own,Garg:2018reu}, meaning that more sophistication may be needed to construct fully realistic models.
The same holds true for the initial conditions of inflation, which often require a certain amount of fine-tuning, depending on the speculative assumptions about the pre-inflationary era.
This is of course undesirable as the entire \textit{raison d'\^etre} of inflation is to resolve the fine-tuning issues related to the initial conditions of big-bang cosmology.
For some authors, these shortcomings of inflation serve as a motivation to consider some of its (arguably less popular) extensions or alternatives, such as as big-bounce models~\cite{Brandenberger:2012zb}, cyclic models~\cite{Khoury:2001wf}, or string gas cosmology~\cite{Battefeld:2005av}.
Most authors, however, adhere to the paradigm of inflationary cosmology, remaining confident that future work in field theory, string theory, and quantum cosmology~\cite{Bojowald:2015iga}, in combination with more data, will resolve the outstanding issues.


\subsection{Hot thermal phase}
\label{subsec:rd}

\begin{figure}

\centering
\includegraphics[width=0.9\textwidth]{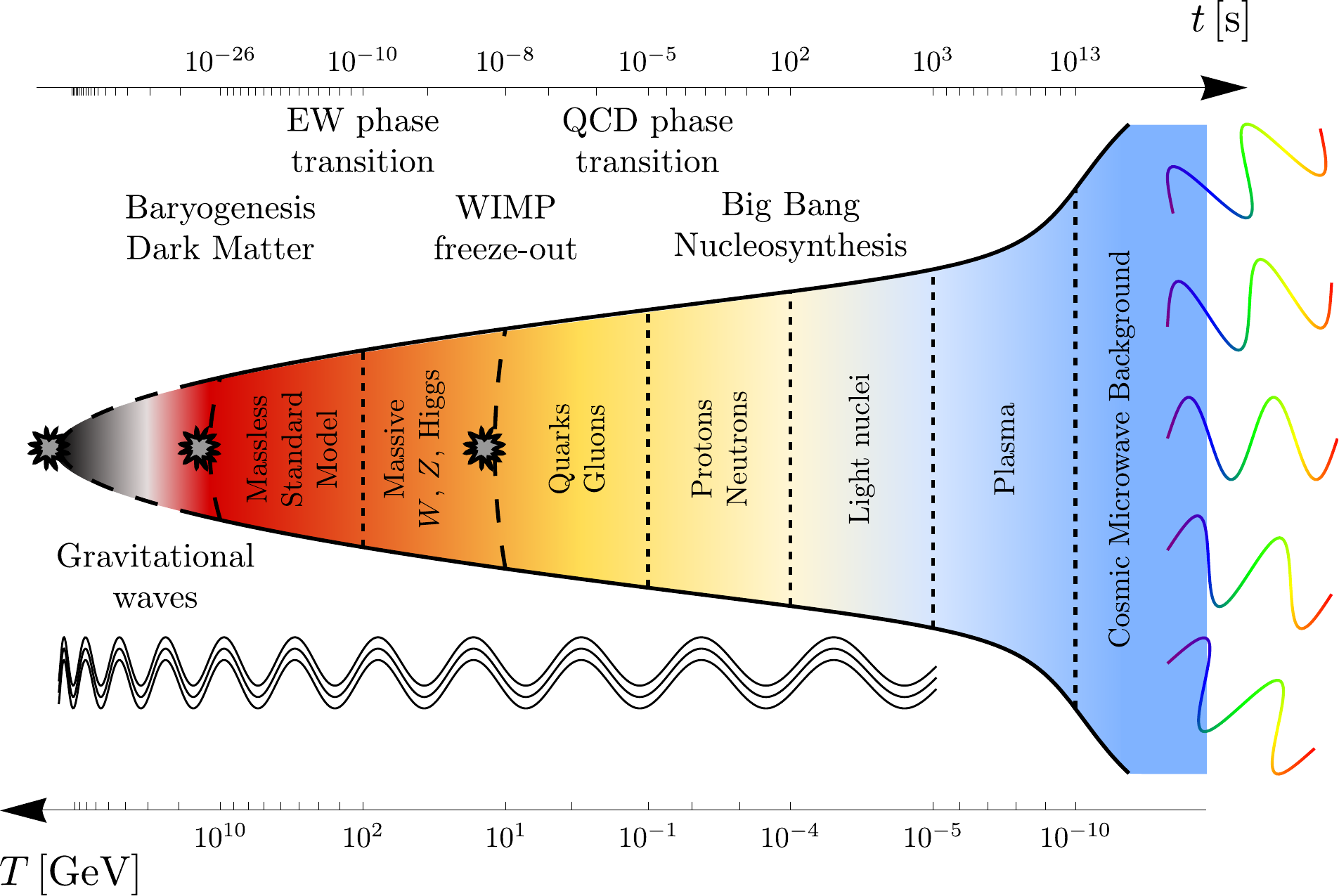} 

\caption{Chronology of the hot big bang, \textit{i.e.}, the radiation-dominated era after inflation and reheating~\cite{Schmitz:2012kaa}.
Three possible values of the maximal temperature in the early Universe are indicated by little gray stars.
Different epochs are labeled by their respective characteristic degrees of freedom and separated by vertical dashed lines.
Gravitational waves produced at very early times can freely propagate through the hot plasma, while photons only decouple from the thermal bath around 380,000 years after the big bang, shortly after matter--radiation equality.
During radiation domination, the scale factor grows like $R \propto t^{1/2}$, while it scales like $R \propto t^{2/3}$ during matter domination, as indicated by the growing width of the black solid envelope.
Image credit: The author.}
\label{fig:history}

\end{figure}

Inflation ends whenever the equation of state of the inflaton field (or fields) no longer supports an accelerating expansion.
This happens, \textit{e.g.}, when the inflaton field picks up a large kinetic energy, \textit{i.e.}, when the so-called slow-roll conditions become violated, or when a critical field value triggers a phase transition in scalar field space.
In the latter case, the phase transition must be sufficiently smooth; a first-order phase transition followed by bubble collisions as in early models of inflation would reintroduce unacceptably large inhomogeneities.
In other words, inflation must allow for a \textit{graceful exit} into the stage of decelerating expansion, an exit that does not spoil the  homogeneity and isotropy established by inflation.
During this exit, the vacuum energy stored in the inflaton field is converted to thermal radiation composed of relativistic particles.
In this sense, assuming that inflation was preceded by a primordial radiation-dominated era at extremely high temperatures, one can say that the decay of the inflaton field after inflation \textit{reheats} the Universe.
However, even independently of any assumptions about the pre-inflationary era, the process of entropy production at the end of inflation is nowadays commonly referred to as \textit{reheating}~\cite{Kofman:1997yn}.
Correspondingly, the temperature of the thermal bath at the end of reheating, when the radiation energy density begins to dominate the cosmic energy budget, is known as the reheating temperature $T_{\rm rh}$.
This temperature marks the onset of the hot big bang and is an important quantity in the description of the early Universe.
At present, its value is only weakly constrained, $T_{\rm rh} \sim \textrm{few} \times 10^{-3} \cdots 10^{15}\,\textrm{GeV}$, which leaves room for countless possible scenarios of the evolution of the early Universe between the end of inflation and BBN~\cite{Allahverdi:2020bys}.
In the following, we will, however, ignore the possibility of a nonstandard expansion history and focus on the most important events during the standard radiation-dominated era after reheating; see Fig.~\ref{fig:history}.


Many models predict a very large reheating temperature, far beyond the energy scales that are accessible in terrestrial collider experiments.
In this sense, the hot thermal plasma in the early Universe represents a unique particle physics laboratory that provides the right environment for particle processes at extremely high energies.
This observation serves as the starting point for the field of particle cosmology~\cite{Mambrini:2021aa}, which applies the methods of theoretical particle physics to the early Universe in order to better understand the phenomenology of the Standard Model as well as of hypothetical new-physics scenarios at very high temperatures.


A central subject of particle cosmology is the evolution of the subatomic forces in the early Universe, which is related to the phenomenon of spontaneous symmetry breaking in the context of gauge theories.
At temperatures of a few 100 GeV, the gauge symmetry governing the interactions of elementary particles is given by the SM gauge group, $G_{\rm SM} = SU(3)_C \times SU(2)_L \times U(1)_Y$, which describes the strong and electroweak gauge interactions in the bath of massless SM particles.
The origin of the SM gauge structure is unknown; but there are good reasons to believe that $G_{\rm SM}$ is in fact just a subgroup of a much larger symmetry group $G_{\rm GUT}$ that unifies all subatomic forces at an energy scale of around $10^{16}\,\textrm{GeV}$, a hypothesis that goes by the name of \textit{grand unification}.
In each grand unified theory, the gauge groups $G_{\rm GUT}$ and $G_{\rm SM}$ are connected by a characteristic pattern of symmetry breaking steps, some of which may take place before inflation (\textit{e.g.}, in order to avoid a monopole problem; see Sec.~\ref{subsec:ics}) and some of which may take place after inflation.
Similarly, the electroweak part of the SM gauge group itself, $SU(2)_L \times U(1)_Y$, becomes spontaneously broken at temperatures of around 160 GeV in consequence of the SM Higgs mechanism.
This mechanism manifests itself in a smooth cosmological phase transition, the electroweak crossover, that yields a nonzero vacuum expectation value of the Higgs field, which in turn induces masses for all SM particles (possibly except for neutrinos whose masses call for new physics). 


After the electroweak phase transition, the temperature keeps decreasing.
Heavy degrees of freedom, such as the top quark or electroweak gauge bosons, are therefore no longer produced in the thermal bath, while the majority of the remaining light degrees of freedom contribute to the so-called quark--gluon plasma, the state of strongly interacting matter in the high-temperature limit of quantum chromodynamics (QCD).
At temperatures of around $160\,\textrm{MeV}$, the plasma of quarks and gluons become confined in color-neutral hadrons.
This process manifests itself again in a smooth phase transition, the QCD crossover or simply quark--hadron phase transition.
Shortly thereafter, baryons and antibaryons begin to decouple, which results in the annihilation of more or less all antibaryons and the freeze-out of a small relic baryon density (see Sec.~\ref{subsec:visible}).
While this process takes place only after the QCD phase transitions, the primordial asymmetry between baryons and antibaryons was presumably already seeded much earlier.
Standard thermal leptogenesis, \textit{e.g.}, one of the most popular baryogenesis scenarios, operates at temperatures of the order of $10^9\,\textrm{GeV}$ or even higher.


Around one second into the radiation-dominated era, the thermal bath consists of relativistic photons, neutrinos, electrons, and positrons as well as nonrelativistic protons and neutrons.
At this point in time, neutrinos participate for the last time in scattering processes mediated by the weak nuclear force, before they decouple and begin to free-stream in the form of the cosmic neutrino background (C$\nu$B).
Like the CMB, the C$\nu$B is a firm prediction of big-bang cosmology.
Because of the elusive nature of neutrinos, it has not yet been observed~\cite{KATRIN:2022kkv}, although experiments such as PTOLEMY~\cite{PTOLEMY:2019hkd} aim at its direct detection in the not-too-distant future.
Shortly after neutrino decoupling, at temperatures around the electron mass scale, electrons and positrons annihilate into photons.
This leaves behind a relic electron density (which is tied to the relic proton density by the requirement that the Universe must remain charge-neutral) and slightly heats up the photon bath.
As a consequence, the CMB is expected to have a slightly larger temperature than the C$\nu$B.
Or conversely, while the CMB temperature today is known to be 2.73 kelvin, the C$\nu$B is predicted to have a temperature of only 1.95 kelvin.


After two minutes or so, primordial nucleosynthesis commences.
The temperature has now dropped to around 100 keV, which is low enough so that the formation of stable deuterium is no longer impeded by photodissociation.
Following the passage of this so-called deuterium bottleneck, a network of nuclear reactions unfolds, which mostly leads to the synthesis of deuterium, helium-3, helium-4, and lithium-7 (see Sec.~\ref{subsec:bbn}).
In particular, all neutrons become bound in atomic nuclei, with the vast majority of neutrons ending up in helium-4.
After around 20 minutes, BBN concludes, which determines the characteristic composition of the plasma in terms of photons, electrons, protons, and light nuclei that it will keep for the entire remainder of the radiation-dominated era.


The end of the radiation-dominated era is reached around 50,000 years after the big bang, when the relic density of dark matter (see Sec.~\ref{subsec:dm}) begins to dominate the energy budget of the Universe.
Dark matter is presumably produced at very high temperatures, maybe already during reheating after inflation.
In a popular class of models, dark matter consists, \textit{e.g.}, of weakly interacting massive particles (WIMPs), particles with a mass of the order of 100 GeV and an interaction strength comparable to the strength of electroweak interactions~\cite{Jungman:1995df}.
Such particles can be thermally produced in the early Universe, before their relic density freezes out at temperatures that are about a factor 30 smaller than their mass (see Fig.~\ref{fig:history}).
This is, however, only one among an almost countless number of possible scenarios for the production of dark matter in the early Universe.
In any case, it is clear that, once the dark-matter particles become nonrelativistic, their energy density decreases because of the cosmic expansion in proportion to $R^{-3}$.
That is, the energy stored in the rest mass of the dark-matter particles remains constant, while the volume that the particles are contained in grows like $R^3$.
This needs to be compared to the dilution of the radiation energy density, which decreases in proportion to $R^{-4}$.
Here, three powers of the scale factor account again for the volume expansion of the Universe, while the fourth power is a consequence of cosmological redshift, \textit{i.e.}, the fact that the wavelength of radiation traveling through the expanding Universe becomes stretched.
Because of these different scaling laws, the energy density of matter necessarily begins to dominate over the energy density of radiation sooner or later.
Given the relic density of dark matter today, which can be measured in CMB observations, one is then able to compute the time of matter--radiation equality in the early Universe, 50,000 years after the big bang, which marks the onset of the matter-dominated era.


Most of the events described above occur before BBN and thus fall into uncharted territory.
In particular, because the early Universe is opaque to photons throughout radiation domination, no signal in the form of electromagnetic radiation will ever reach us from this era.
Therefore, in order to push the frontier of our knowledge further back in time, an important line of attack necessarily consists of indirect probes, \textit{i.e.}, astrophysical and cosmological observations of the late Universe after photon decoupling in combination with terrestrial experiments.
However, alongside these indirect approaches, there is also one direct messenger from the early Universe that can give us a peek behind the veil of the CMB: primordial gravitational waves~\cite{Caprini:2018mtu}, which travel more or less freely through the early Universe after their production.
Many processes in the early Universe can give rise to a stochastic background of primordial gravitational waves, ranging from inflation and reheating over first-order cosmological phase transitions to topological defects such as domain walls and cosmic strings.
The detection of a such a background would undoubtedly mark a major milestone in the still young field of GW astronomy.


In fact, a new low-frequency signal has recently shown up in the data of pulsar timing arrays (PTAs), which monitor networks of pulsars across the Milky Way in order to search for nanohertz gravitational waves~\cite{Taylor:2021aa}.
All big PTA collaborations, NANOGrav, PPTA, EPTA, and IPTA, have now found strong evidence for this signal~\cite{NANOGrav:2020bcs,Goncharov:2021oub,Chen:2021rqp,Antoniadis:2022pcn}, albeit it is not yet clear whether it really possesses all the properties of a genuine GW signal.
More PTA data and analyses in the coming years will help clarify the situation and bring us closer to the first detection of a stochastic GW background.
It is expected that this background will receive large contributions from inspiraling supermassive black-hole binaries at the center of merging galaxies.
With a bit of luck, however, the data may also contain evidence for the presence of gravitational waves from the early Universe.


\subsection{Structure formation}
\label{subsec:structure}


During the radiation-dominated era, the shape of the gravitational potential is controlled by density perturbations in the radiation bath.
This prevents the efficient growth of structure at early times, as the pressure of the baryon--photon fluid counteracts the gravitational pull towards overdense regions.
On top of that, the fast cosmic expansion during radiation domination constantly drags all particles apart, which causes the gravitational potential to actually decay.
The onset of efficient structure formation is therefore delayed to the time of matter--radiation equality, when the CDM density perturbations begin to take control.
From this point on, the gravitational potential remains stable, which goes hand in hand with a more rapid accumulation of dark-matter particles in gravitational potential wells.
Dark matter thus already begins to collapse under the influence of gravity, while the evolution of the baryon--photon fluid is still governed by the characteristic sound waves that later become imprinted in the CMB.
Indeed, the growth of baryonic density perturbations only speeds up after photon decoupling, 380,000 years after the big bang, when the baryons begin to fall towards the pre-existing dark-matter overdensities.
Remarkably enough, the sound-wave pattern in the baryon density partially survives this process, which eventually leads to the BAO signature in the galaxy power spectrum at later times.


After photon decoupling, the CMB radiation becomes redshifted by the cosmic expansion, so that it quickly leaves the visible part of the electromagnetic spectrum.
At this point, the Universe no longer contains any visible light sources, which marks the beginning of the so-called dark ages.
During this epoch, most of the baryonic matter resides in neutral hydrogen, which occasionally absorbs CMB photons in 21-cm spin-flip transitions~\cite{Furlanetto:2006jb}.
At the same time, the inhomogeneities in the baryon density continue to increase, such that structure formation transitions from the linear to the nonlinear regime on small scales (in a hierarchical bottom-up fashion; see Sec.~\ref{subsec:dm}). 
Roughly hundred million years into the dark ages, this leads to the formation of the first generation of stars (\textit{i.e.}, population III stars), followed by the formation of the first early galaxies and quasars (\textit{i.e.}, active galaxies with supermassive black holes at their centers) over the subsequent few hundred million years.


The neutral hydrogen partially absorbs and re-emits the ultraviolet Lyman-alpha radiation from the first stars, before it is heated by luminous X-ray sources, such as active galactic nuclei or the hot interstellar medium.
This epoch is known as cosmic dawn, which results in a second characteristic 21-cm absorption feature in the CMB background.
In 2018, the EDGES collaboration announced the first detection of this signal with an unexpectedly large amplitude~\cite{Bowman:2018yin}; however, since then, no other experiment has been able to confirm this claim~\cite{Singh:2021mxo}.
Future radio observations by LOFAR~\cite{vanHaarlem:2013dsa,Shimwell:2022oot}\,/\,NenuFAR~\cite{Mertens:2021apf} and the Square Kilometre Array (SKA)~\cite{Mellema:2012ht} will be crucial to clarify the case.
Meanwhile, the epoch of cosmic dawn will also be probed by the James Webb Space Telescope, which was launched in 2021 and is expected to collect light from the first stars and galaxies~\cite{Gardner:2006ky}. 


Another important effect of the first stellar light in the Universe is that it gradually ionizes the neutral hydrogen gas.
This process is referred to as reionization~\cite{Barkana:2000fd}, lasts from around 250 million years to around one billion years after the big bang, and is powered by the energetic ultraviolet radiation emitted by the first stars, galaxies, and quasars, which separates again electrons from protons and hence reverts the outcome of recombination.
Direct evidence for the epoch of reionization is, \textit{e.g.}, provided by quasar absorption spectra.
High-redshift quasars feature broadened and redshifted Lyman-alpha absorption lines of neutral hydrogen~\cite{SDSS:2001tew,Fan:2005es}, known as Gunn--Peterson troughs~\cite{Gunn:1965hd}, while the spectra of quasars located at lower redshifts exhibit no such troughs.
Based on these observations, one concludes that the intergalactic medium must have been neutral at early times, before it became fully ionized towards the end of the first billion years.
After reionization, neutral hydrogen only remains in isolated gas clouds, which results in a sequence of individual Lyman-alpha absorption lines in the spectra of distant quasars.
This so-called Lyman-alpha forest can be used as a tracer of the large-scale structure of the Universe and hence as a tool to constrain cosmological parameters~\cite{Seljak:2006bg}.
A closely related probe of the large-scale structure consists of hydrogen intensity mapping, \textit{i.e.}, the measurement of 21-cm emission from unresolved clouds of neutral hydrogen~\cite{Pritchard:2011xb}.
The promise of this technique is that it will eventually enable us to map the three-dimensional matter distribution across vast spatial volumes.
In a first step towards this goal, the CHIME experiment recently reported the detection of 21-cm emission from select galaxies and quasars~\cite{CHIME:2022kvg}; next, it aims at measuring the BAO signature in the auto correlation power spectrum of 21-cm emission.


Structure formation continues throughout and after reionization, which gradually results in what is known as the \textit{cosmic web}, the familiar distribution of matter that shapes the appearance of the Universe up to this day.
The cosmic web consists of massive dark-matter halos, connected to each other by filamentary structures and surrounded by vast underdense voids.
In the cosmic web, the baryonic matter traces the distribution of cold dark matter, which results in the formation of galaxies in overdense regions. 
Over time, these galaxies assemble in galaxy clusters, with the most massive superclusters forming at the center of dense dark-matter halos.
In this process, the baryonic matter can condense into complex structures even on very small scales, because, unlike dark matter, it is able to cool and loose angular momentum via the emission of electromagnetic radiation.

\begin{figure}[t]

\centering
\includegraphics[width=\textwidth]{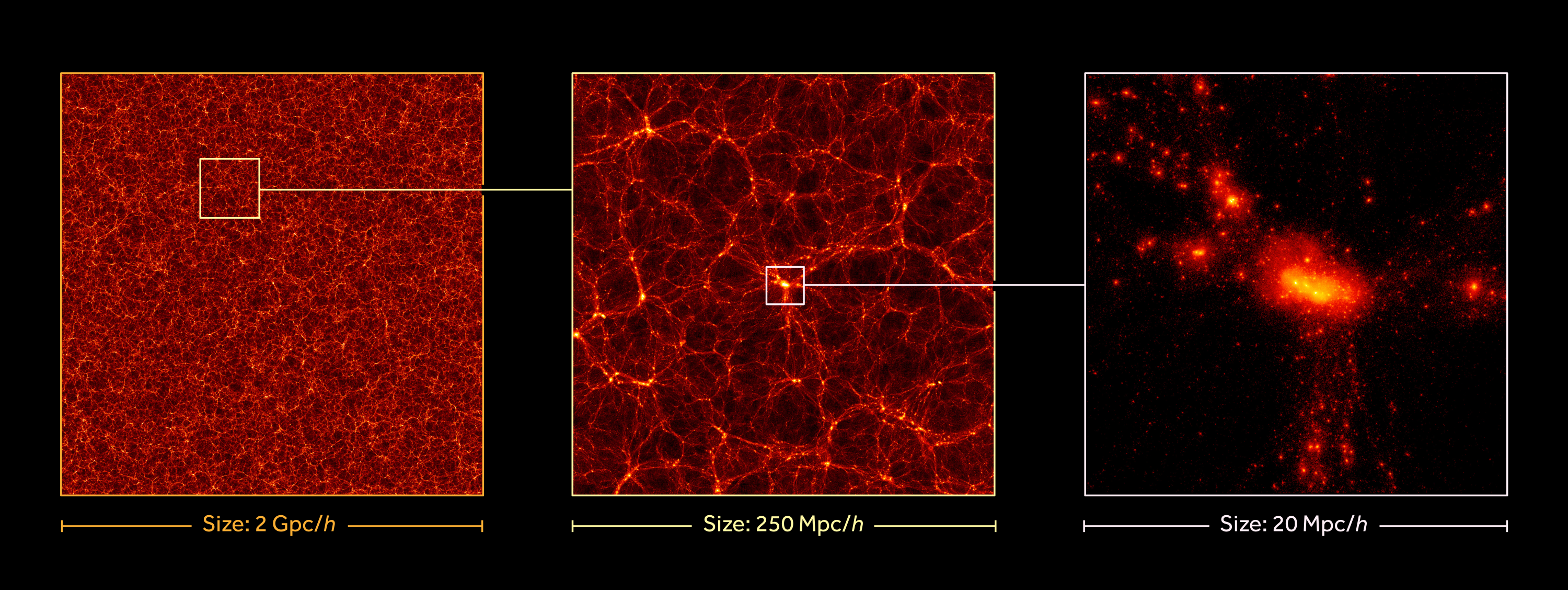} 

\caption{Visualization of the AbacusSummit simulation~\cite{Maksimova:2021ynf}, a state-of-the-art cosmological N-body simulation of roughly 330 billion test particles inside a volume of $\left(2\,\textrm{Gpc}/h\right)^3$, where $h \approx 0.7$ is a dimensionless measure for the Hubble constant $H_0$.
On gigaparsec scales (left), the matter distribution is homogeneous and isotropic in accord with the cosmological principle; on scales of around 100 megaparsec (middle), the cosmic web consisting of halos, filaments, and voids becomes apparent; on scales of around 10 megaparsec (right), individual halos and galaxy clusters are resolved.
Image credit: The AbacusSummit Team.}
\label{fig:nbody}

\end{figure}

The dynamics and evolution of the cosmic web is a highly nonlinear process that is best studied in numerical N-body simulations (see Fig.~\ref{fig:nbody}), an important tool of modern cosmology.
State-of-the-art N-body simulations are very successful in describing the evolution of structure on large length scales.
On shorter length scales, however, they currently face a number of challenges that demand further scrutiny~\cite{Bullock:2017xww}.
One of these small-scale problems consists, \textit{e.g.}, in the observation that most dwarf galaxies appear to host a dark-matter core of roughly constant density at their center, while simulations based on the $\Lambda$CDM model rather indicate a cuspy density profile.
Other problems, next to this cusp--core problem, are known as the missing-satellites problem or the too-big-to-fail problem.
The origin of these small-scale problems might be related to the complicated baryonic feedback on structure formation on small scales, which is currently not well understood, or point to nonstandard properties of dark matter, such as weak dark-matter self-interactions or a warm dark-matter component.
Progress at this frontier will require improved theoretical and numerical modelling as well as more astronomical data, such as, \textit{e.g.}, the Legacy Survey of Space and Time (LSST) conducted by the Vera C.\ Rubin Observatory~\cite{LSSTScience:2009jmu}.
 

Our own galaxy, the Milky Way, forms around five billion years after the big bang, followed by the solar system roughly four billion years later.
Another billion years later, around ten billion years after the big bang, dark energy begins to dominate the cosmic energy budget, which marks the onset of the late-time accelerated expansion (see Sec.~\ref{subsec:de}).
This transition from matter domination to dark-energy domination is a consequence of the fact that the energy density of matter steadily decreases because of the volume expansion, $\rho_m \propto R^{-3}$, while the energy density of dark energy remains constant, $\rho_\Lambda \propto R^0$.
The cosmic expansion thus constantly increases the absolute amount of dark energy, which may appear counterintuitive from the perspective of classical mechanics, but which is indeed a perfectly viable possibility in the context of general relativity in an expanding Universe.
An important effect of dark energy on structure formation is that it causes the gravitational potential to decay again.
In the late Universe, the growth of structure because of gravitational instability can no longer compete with the accelerated expansion, such that wells and hills in the gravitational potential begin to decay.
CMB photons traveling through this decaying potential landscape will therefore pick up small additional gravitational red- and blueshifts.
This phenomenon is known as the late-time integrated Sachs--Wolfe effect~\cite{Sachs:1967er}, or Rees--Sciama effect~\cite{Rees:1968zza}, and has been observed in the cross correlation of the CMB with the large-scale structure~\cite{Fosalba:2003ge,SDSS:2003lnz}. 


In the coming billion years, our Universe will continue to appear similar to its present state, albeit the accelerated expansion will continue to push more and more galaxies beyond our cosmic horizon.
Thus, assuming that the accelerated expansion continues far into the future, the observable Universe will eventually only contain the gravitationally bound group of galaxies in our immediate cosmic neighbourhood; and these galaxies will gradually grow dimmer over time as the fuel for the formation of new stars becomes depleted.
This extrapolation, however, depends on our assumptions about the nature of dark energy and therefore needs to be treated with caution.
Statements about the ultimate fate of the Universe are even more speculative, whereby we have come full circle and returned to the starting point of our chronology of the Universe in this section.


\section{Status and prospects}


In the course of the hundred years since Einstein and Hubble, cosmology has advanced from a subdiscipline of astronomy to a cutting-edge research field in its own right, closely interwoven with other strands of basic science.
Today, cosmology is an interdisciplinary and international endeavor that epitomizes the human quest for knowledge, humanity's desire to better understand the cosmos and its place in it.
To this end, cosmology brings together insights from many neighbouring fields, ranging from string theory, field theory and high-energy physics over neutrino physics, astroparticle physics, and astronomy to nuclear, atomic, and gravitational physics.
Modern cosmological observations and simulations are in particular a prime incarnation of big-data science and benefit as such from the steady advances in computing power and new algorithms, \textit{e.g.}, based on machine learning~\cite{Ntampaka:2019udw,Villaescusa-Navarro:2020rxg,Villaescusa-Navarro:2022qqq}. 
In addition, the diverse observational program of modern cosmology is a shining example of collaboration science that capitalizes on community practices such as queue observing and open-access archives. 


The progress in the field over the last century and especially over the last decades has been impressive. 
Based on several major breakthrough detections, including the Hubble expansion, the CMB and its anisotropies as well as dark energy, cosmology now possesses a rigorously formulated standard model, the $\Lambda$CDM concordance model, which manages to explain the bulk of all cosmological data with great precision in terms of just a few basic building blocks: ordinary baryonic matter, cold nonbaryonic dark matter, and a cosmological constant.
We now understand that matter density perturbations, probably seeded by an early stage of cosmic inflation before the hot big bang, gave rise to the large-scale structure of the Universe, while the cosmological constant completes the cosmic energy budget required for a flat Universe, which causes the present-day expansion to accelerate.
Since the COBE measurement of the CMB blackbody spectrum in the early 1990s, we have moreover entered the era of precision cosmology~\cite{Turner:2022gvw}.
A large number of cosmological parameters are now known with percent-level precision, which is a testament to the long way that modern cosmology has come since the days of Hubble, whose determination of the Hubble constant, $H_0 \approx 500\,\textrm{km}/\textrm{s}/\textrm{Mpc}$, was not even of the correct order of magnitude.


Despite its overall success, the $\Lambda$CDM model as well as the precise determination of its parameters leave us with profound questions that remain to be answered. 
First of all, it is remarkable that all of the basic $\Lambda$CDM building blocks require new physics beyond the Standard Model of particle physics:
The Standard Model does not contain a viable particle candidate for dark matter; naive QFT estimates of the energy density of the vacuum exceed the observed value of the cosmological constant by an embarrassingly large amount; and the cosmic baryon--antibaryon asymmetry calls for an exotic nonequilibrium process in the early Universe.
At the same time, big-bang cosmology, while successful in explaining the primordial abundances of the light elements as well as the properties of the CMB, lacks a satisfying explanation of its initial conditions.
The current paradigm for the dynamics underlying these initial conditions is cosmic inflation, which represents yet another phenomenon that cannot be explained in terms of SM physics in combination with Einstein gravity.  
Together, these open questions are key drivers of the research program of particle physics and cosmology in the 21$^{\rm st}$ century.


More challenges arise from the precise measurement of cosmological parameters, which put the $\Lambda$CDM framework itself to the test.
Now, in the era of precision cosmology, different routes to measuring the same observable can especially be used to set up powerful consistency checks that may reveal even slight discrepancies.
Indeed, in the recent past, various such tensions have emerged, some more subtle and others more severe~\cite{Perivolaropoulos:2021jda}.
The list of tensions, challenges, and oddities includes (i) various CMB anomalies~\cite{Schwarz:2015cma}, such as a lack of sizable temperature correlations on large angular scales or gravitational-lensing effects above the expected level; (ii) the small-scale problems of structure formation (see Sec.~\ref{subsec:structure}), such as the cusp--core or missing-satellites problems; (iii) observations challenging the cosmological principle, such as dipole measurements, \textit{e.g.}, in the distribution of quasars~\cite{Secrest:2020has}, that do not match the properties of the CMB dipole; and several others.
On top of that, there is an on-going debate related to different measurements of the clustering amplitude $\sigma_8$, an observable that characterizes the size of matter density fluctuations in the late Universe (in the linear-theory approximation) on length scales of the order of $8\,\textrm{Mpc}/h$, \textit{i.e.}, on galaxy-cluster scales (see Fig.~\ref{fig:nbody}).
Some weak-lensing and galaxy-clustering surveys find $\sigma_8$ values below the value inferred by the PLANCK observations of the CMB, at a statistical significance of up to roughly $3\,\sigma$~\cite{Heymans:2020gsg,KiDS:2020suj}, while other surveys find values closer to the CMB value~\cite{DES:2021wwk}.
It therefore remains to be seen whether the $\sigma_8$ discrepancy will eventually reveal a crack in the $\Lambda$CMB paradigm or not.


The elephant in the room, however, dwarfing all of the anomalies listed above, is the so-called $H_0$ or Hubble tension~\cite{Verde:2019ivm,DiValentino:2021izs}, a mismatch between different determinations of the Hubble constant that has recently reached a statistical significance of $5\,\sigma$~\cite{Riess:2021jrx} and which has even led some commentators to proclaim a \textit{crisis in cosmology}.
Of course, such a proclamation is overly dramatic and in a sense also misleading: 
If anything, the Hubble tension illustrates that cosmology as a scientific discipline is in fact in a very healthy state.
It has now become a mature discipline of precision science, where a steady influx of new data enables one to make real and meaningful progress.
In the case of the Hubble tension, this state of affairs is reflected in increasingly precise determinations of the Hubble constant.
On the one hand, the ``global'' measurement of $H_0$ based on CMB observations by the PLANCK satellite, assuming a standard $\Lambda$CDM cosmology, yields a value of $H_0 = 67.4 \pm 0.5\,\textrm{km}/\textrm{s}/\textrm{Mpc}$~\cite{Planck:2018vyg}.
On the other hand, recent ``local'' measurements at low redshift based on the observation of Cepheid variables and Type Ia supernovae  (\textit{i.e.}, the \textit{cosmic distance ladder}) consistently find larger values, $H_0 = 73.0 \pm 1.0\,\textrm{km}/\textrm{s}/\textrm{Mpc}$~\cite{Riess:2021jrx}.
The origin of this discrepancy is unknown, which currently fuels intense research efforts in particle physics and cosmology aimed at identifying possible solutions.
Next to systematic effects, which may account for parts of the tension, various possibilities are presently scrutinized, including departures from $\Lambda$CDM in the early Universe, departures from $\Lambda$CDM in the late Universe, as well as combinations thereof~\cite{Schoneberg:2021qvd}.
Many proposed solutions also promise other observational signatures that may eventually enable one to identify the correct solution.
However, when and in which specific way this is going to happen is impossible to predict at the moment.


In any case, it is clear that cosmology is facing a bright future in the coming years and decades, regardless of whether the Hubble tension will still stay with us for a bit longer or not.
A comprehensive outlook is beyond the scope of this essay; however, a few exciting prospects deserve to be highlighted.
On the observational side, a wealth of data from next-generation space telescopes is awaiting us, including NASA's James Webb~\cite{Gardner:2006ky} and Nancy Grace Roman~\cite{Eifler:2020vvg} as well as ESA's Euclid~\cite{Amendola:2016saw} and Athena~\cite{Nandra:2013jka} space missions.
These space telescopes will be complemented by new optical reflecting telescopes on the ground, such as the Vera C.\ Rubin Observatory~\cite{LSSTScience:2009jmu}, the Thirty Meter Telescope~\cite{TMT:2015pvw}, and the Extremely Large Telescope, alongside the next generation of radio and gamma-ray telescopes, SKA~\cite{Mellema:2012ht} and the imaging atmospheric Cherenkov Telescope Array (CTA)~\cite{CTAConsortium:2017dvg}, as well as a suite of new space- and ground-based CMB experiments (see Sec.~\ref{subsec:cmb}).
These new telescopes and observatories will enable a revolutionary research program of extraordinary depth and breadth, which will take proven observational methods (\textit{e.g.}, galaxy surveys, weak gravitational lensing, supernova cosmology) to new heights and establish novel techniques that have currently not yet reached their full potential (\textit{e.g.}, cosmic chronometers, gamma-ray bursts~\cite{Biteau:2022dtt}, 21-cm cosmology) as new tools in the cosmologist's toolbox~\cite{Moresco:2022phi,Wu:2022dgy}.


Moreover, we can expect groundbreaking developments at the multi-messenger frontier, \textit{i.e.}, in the fields of neutrino and GW cosmology.
In particular, the third generation of ground-based GW interferometers, Cosmic Explorer~\cite{Reitze:2019iox} and Einstein Telescope~\cite{Maggiore:2019uih}, as well as the first GW space missions, the Laser Interferometer Space Antenna (LISA)~\cite{LISA:2017pwj}, Taiji~\cite{Hu:2017mde}, and TianQin~\cite{TianQin:2015yph}, promise to open new windows onto the Universe.
GW events can, \textit{e.g.}, be used as \textit{standard sirens}~\cite{LIGOScientific:2017adf}, in analogy to type-Ia supernova \textit{standard candles}, in order to measure cosmological parameters. 
Furthermore, the detection of a cosmological GW background would grant us access to processes in the very early Universe that are currently hidden behind the veil of the CMB.
In fact, the current generation of PTA observations may actually be on the brink of detecting a GW background at nanohertz frequencies (see Sec.~\ref{subsec:rd}).
While this signal is expected to be dominated by the astrophysical foreground from supermassive black-hole binaries, there is a slight chance that it may also contain a cosmological component.


Finally, the rich observational program in the coming years will also stimulate and profit from theoretical progress, including the development of new models, methods, and algorithms. 
We thus find ourselves on the eve of a golden age of cosmology that will likely shape our understanding of the cosmos in ways that we can barely imagine at present.
Along the way, we may gradually encounter solutions to the various challenges and tensions of the $\Lambda$CDM model, which may or may not usher in the next paradigm shift in cosmology.
Either way, a fascinating journey lies ahead of us that is set to resolve various deep questions and give rise to new ones.
The first century of modern cosmology has been a true adventure, the second century will be even more exciting.


\bibliographystyle{JHEP}
\bibliography{arxiv_1}{}


\end{document}